\documentclass[aps,prd,twocolumn,showpacs,groupedaddress,nofootinbib,longbibliography]{revtex4-2}
\usepackage{stmaryrd}

\usepackage{times}
\usepackage{amsfonts}
\usepackage{graphicx}
\usepackage{amsmath}
\usepackage{amssymb}
\usepackage{color}
\usepackage{subfigure}
\usepackage{natbib}
\usepackage{tensor}
\usepackage{accents}
\usepackage[colorlinks=true, citecolor=blue]{hyperref}
\usepackage{xspace}
\usepackage{xfrac}

\def\t{\theta}
\def\be{\begin{equation}}
\def\ee{\end{equation}}
\def\ra{\rangle}

\def\diff{\text{\rm Diff}^+\!(S^1)}
\def\adiff{\mathfrak{diff}(S^1)}
\def\psl{\text{\rm PSL}(2, \bb R)}

\def\vira{\text{\sl Vira}}
\def\avira{\mathfrak{vira}}
\def\dvira{\mathfrak{vira}^*}

\def\ads{\text{\rm AdS}_3}

\newcommand{\ca}[1]{\mathcal{#1}}
\newcommand{\bb}[1]{\mathbb{#1}}
\newcommand{\fr}[1]{\mathfrak{#1}}

\newcommand{\wt}[1]{\widetilde{#1}}
\newcommand{\wh}[1]{\widehat{#1}}

\begin{document}

\title{Causal diamonds in 2+1 dimensional quantum gravity }

\author{Rodrigo Andrade e Silva}
\email{rasilva@umd.edu} 
\author{Ted Jacobson}
\email{jacobson@umd.edu}
\affiliation{Maryland Center for Fundamental Physics, 
University of Maryland, College Park, MD 20742}

\begin{abstract}
We develop the reduced phase space quantization of causal diamonds in pure 2+1 dimensional gravity with a non-positive cosmological constant. The system is defined as the domain of dependence of a topological disc with fixed boundary  metric. By solving the initial value constraints in a constant-mean-curvature time gauge and removing all the spatial gauge redundancy, we find that the phase space is the cotangent bundle of $\diff/\psl$. To quantize this  phase space we apply Isham's group-theoretic quantization scheme, with respect to a BMS$_3$ group, and find that the quantum theory can be realized by wavefunctions on some coadjoint orbit of the Virasoro group, with labels in irreducible unitary representations of the corresponding little group. We find that the twist of the diamond boundary loop is quantized
in integer or half-integer multiples of
the ratio of the 
Planck length to 
the boundary length.
\end{abstract}

\maketitle

\section{Introduction}

Among the many challenges to understanding nonperturbative 
quantum gravity are that 
standard canonical quantization is inapplicable 
due to the nonlinearity of the phase space,
that local observables are not available,
and that general
relativity in four or more spacetime dimensions is (likely) 
not an ultraviolet-complete quantum field theory.
On top of those is the obstacle of removing the diffeomorphism
gauge redundancy (a.k.a. ``coordinate freedom"), and 
the fact that spacetime diffeomorphisms include deformations
in timelike directions, making time evolution
a gauge transformation, which leads to the vexing 
``problem of time" \cite{isham1993canonical, kuchavr2011time, anderson2012problem}.
To make progress it is worthwhile to consider 
simplified settings, and over the past several decades
much work of that nature has been done.
Here we consider a new such setting, in which 
all of the above-mentioned challenges can be met,
namely, causal diamonds in 2+1 dimensional general relativity
with a negative cosmological constant.

\begin{figure}
\centering
\includegraphics[scale = 0.4]{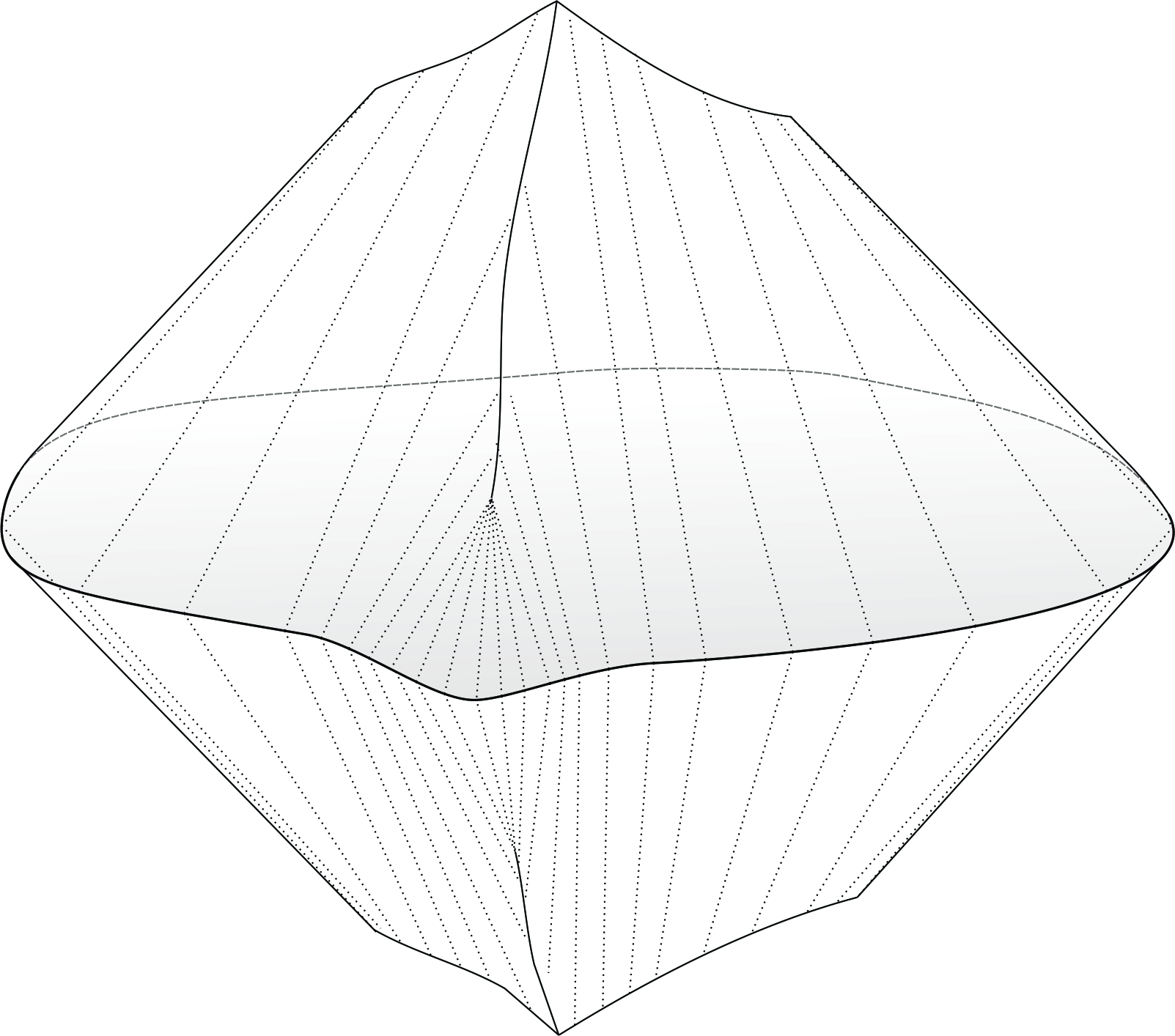}
\caption{A generic classical state corresponds to a causal diamond in $\ads$ (or in Mink${}_3$ if $\Lambda=0$)
with boundary length $\ell$. Note that in general the 
Cauchy horizon is not smooth since the null generators exit at caustics.}
\label{fig:diamond}
\end{figure}

By a 2+1 causal diamond we mean
the domain of dependence of a spacelike topological disc with fixed boundary metric. 
To quantize the system we employ the reduced phase space approach, in which we first impose 
all the initial value constraints and remove the gauge ambiguities at the classical level, and then proceed with the quantization. Since there are no local degrees of freedom in 2+1 gravity, and 
we choose the topology of the spatial slices to be 
that of
a disc, the classical states (solutions to the Einstein equation, up to gauge transformations)
can only correspond to all possible 
shapes of causal diamonds, 
with boundary length $\ell$ determined by the fixed boundary metric,
embedded in Anti-de Sitter space ($\ads$)
if $\Lambda <0$ or in Minkowski space (Mink${}_3$) if $\Lambda=0$ 
(see Figure \ref{fig:diamond}). 
We find that the corresponding
phase space is 
the cotangent bundle 
$T^*\cal Q$ of a configuration space
${\cal Q} = \diff/\psl$
that is the quotient of 
the infinite dimensional 
group of orientation preserving smooth 
maps of the boundary loop into itself, by
the projective special linear group in two real dimensions
(which is the finite dimensional subgroup of 
$\diff$ induced by conformal isometries\footnote{In this paper a
{\it conformal transformation} 
acts on tensors as multiplication by a positive function followed with the push-forward by a diffeomorphism. Metrics related by such a transformation are said to be {\it conformally equivalent}; and a transformation that leaves the metric invariant is called a {\it conformal isometry}.}
of the unit flat disc). 
Similar 2+1 gravity systems have been considered in the literature, such as spacetimes with closed spatial slices (where the reduced phase space is finite-dimensional) \cite{
witten19882+,witten1989topology,
moncrief1989reduction,moncrief1990solvable,fischer1997hamiltonian,
ashtekar19892+,
hosoya19902+,
 Carlip:1998uc, carlip2005quantum},
spacetimes with finite timelike boundary~\cite{kraus20213d,adami2020sliding,ebert2022field}, and asymptotically $\ads$ spacetimes\cite{brown1986central,freidel20042+,carlip2005conformal,
witten2007three,maloney2010quantum,
Scarinci:2011np,kim2015canonical, cotler2019theory}. 
The causal diamonds provide a novel, quasi-local
system of quantum gravity in
globally-hyperbolic spacetimes 
that, while simple enough to be exactly solvable
classically, 
has an infinite-dimensional reduced phase space of ``boundary gravitons''.

In this paper we describe the classical reduction process and explain how to quantize the resulting  phase space using Isham's 
scheme~\cite{isham1984topological,isham1989canonical} in which the quantization is designed to preserve 
a group of symplectic (a.k.a.\ canonical) transformations of the phase space. This ``quantization group" in our
case is the three dimensional
Bondi-Metzner-Sachs group
BMS${}_3$.
We discuss the 
representation theory of the algebra 
of quantum observables, and deduce 
that the twist of the diamond boundary loop---which is proportional to the 
spin of the diamond---is quantized in terms of the ratio of the Planck length to the boundary length.
This paper is a brief summary of 
some aspects of our study, 
the full details of which will appear in \cite{rodrigo2022}.

\section{Classical}

In the Arnowitt-Deser-Misner (ADM) formulation of general relativity~\cite{arnowitt2008republication}, the phase space 
before reduction is described by Riemannian metrics $h_{ab}$ and conjugate momenta $\pi^{ab}$
$=\sqrt{h}(K^{ab}-K h^{ab})$, where $K_{ab}$ is
the 
extrinsic curvature on 
an initial value spatial surface
(Cauchy slice), here
assumed
to have the topology of a disc $D$.\footnote{We adopt units with 
$c = 16\pi G = 1$.}
We shall restrict to metrics that induce a fixed metric on the boundary, $h|_{\partial D} = \gamma$. 
Note however that the total length $\ell$
of the boundary loop is the only gauge invariant attribute of the boundary geometry that is fixed by this condition. 
The maximal development of any data $(h, \pi)$ that satisfy the initial value constraints of general relativity defines a {\sl causal diamond}. 

A natural choice of intrinsic time function 
$\tau$ is given by 
(minus) the mean extrinsic curvature on the leaves of a foliation of the diamond by
constant-mean-curvature (CMC) Cauchy surfaces,  
$\tau = - K^{ab}h_{ab}$. 
The nonpositive cosmological constant $\Lambda\le0$ ensures that, as $\tau$ ranges from $-\infty$ to $+\infty$, the CMC surfaces foliate the diamond~\cite{bartnik1988regularity,bartnik1988remarks, brill1976isolated, gerhardt1983h}.  This gauge-fixing of time also confers great simplification to the Lichnerowicz method 
\cite{lichnerowicz1944integration}
of solving the Einstein constraint equations~\cite{york1972role,moncrief1989reduction}, 
which consist of a scalar constraint
and a vector constraint.
In this method, 
we start with ``seed data"  $(h_{ab}, \pi^{ab})$ 
on a CMC slice with a given value of $\tau$,
satisfying the
boundary condition on $h_{ab}$ and the
vector constraint $\nabla_a\sigma^{ab} = 0$, 
where $\sigma^{ab} := K^{ab} + \frac{1}{2}\tau h^{ab}$ 
is the traceless part of $K^{ab}$ and $\nabla_a$ is the covariant derivative
determined by $h_{ab}$.
Then, by means of a Weyl-transformation, we use this
seed data to
 generate initial data $(\wt h_{ab}, \wt \pi^{ab})$ that satisfy both the vector and the scalar constraints. The new data, defined by $\wt h_{ab} = e^\phi h_{ab}$, $\wt \sigma^{ab} = e^{-2\phi} \sigma^{ab}$ and $\wt \tau = \tau$, 
continue to satisfy the vector constraint
(for any $\phi$), 
satisfy  the boundary condition
iff $\phi|_{\partial D}=0$, 
and 
satisfy
the scalar constraint iff $\phi$ satisfies the  (two-dimensional) Lichnerowicz equation
\be
\nabla^2 \phi - R_{(h)} + e^{-\phi} \sigma^{ab} \sigma_{ab} - e^\phi \chi = 0,
\ee
where $R_{(h)}$ is the scalar curvature of the metric $h_{ab}$ and $\chi = -2\Lambda +\tau^2/2$. The fact that $\chi \ge 0$ ensures that this equation always has a unique solution for $\phi$
given a boundary condition~\cite{o1973existence,rodrigo2022}.

Since
any element in the family of Weyl-deformed data, $(e^\lambda h_{ab}, e^{-2\lambda} \sigma^{ab}, \tau)$,
leads
to the same solution 
$(\wt h_{ab}, \wt \pi^{ab})$ 
of the
initial value 
problem,
the constraint surface on the phase space can be identified with the set of
equivalence classes $[(h_{ab}, \sigma^{ab}) \sim (e^\lambda h_{ab}, e^{-2\lambda} \sigma^{ab})]$. Spatial diffeomorphisms that act trivially at the boundary,
and only those,  correspond to gauge transformations~\cite{rodrigo2022}, hence the reduced phase space (i.e., the space of physically inequivalent solutions to the equations of motion) can be 
identified as the set of equivalence classes
of seed data
\be
[(h_{ab}, \sigma^{ab}) \sim (\Psi_* e^\lambda h_{ab}, \Psi_* e^{-2\lambda} \sigma^{ab})]
\ee
where $\Psi$ is a boundary-trivial diffeomorphism on $D$ (and $\Psi_*$ is the push-forward) and $\lambda$ is a function on $D$ vanishing at the boundary \cite{rodrigo2022}. 
This happens to be the cotangent bundle $T^*\ca Q$ of the 
space $\ca Q$ of metrics on the disc with fixed induced 
boundary metric,
modulo diffeomorphisms and Weyl transformations that are trivial on the boundary;
and, as one might expect, the symplectic structure is the natural one on the cotangent bundle. 
In fact, $\ca Q$ is the homogeneous space $\diff/\psl$,\footnote{In brief, $\diff$ 
(orientation preserving diffeomorphisms of the boundary loop, 
acting together with the corresponding Weyl transformation that preserves the boundary metric) acts transitively 
on $\ca Q$ (since all metrics on a disc are equivalent under conformal transformations that are allowed to act non-trivially at the boundary).
The subgroup that leaves invariant each point of $\ca Q$, e.g., the (equivalence class of the) Euclidean round disc, is 
$\psl$. Therefore 
$\ca Q=\diff/\psl$.}
and thus the reduced phase space is
$\wt{\ca P} = T^*[\diff/\psl]$.
This is the first of our main results.

There is another approach to the phase space reduction based on a suitable change of coordinates from ADM variables to ``conformal coordinates'', which exploits 
 the fact that all metrics on a disc are conformally equivalent. This alternate approach provides an explicit projection map from the concrete geometrical ADM variables to abstract variables describing $\wt{\ca P}$~\cite{rodrigo2022}. It is useful for several constructions, and relevant when physically interpreting the meaning of observables in the quantum theory, but we 
 postpone its discussion to Sec.~\ref{confcoord} since it is not required for the quantization procedure.

The Hamiltonian 
generating evolution in $\tau$
on the reduced phase space can be obtained by starting with the
Einstein-Hilbert action in the ADM form and then re-expressing it in terms of variables on the reduced phase space. 
The action $S[{\cal C}]$ along a
curve ${\cal C}$
in the (constrained) ADM phase space is
\be\label{action}
S[{\cal C}] = \int_{\cal C} dt  \int_D d^2\!x\, \pi^{ab} \dot h_{ab}  =
\int_{\wt{\cal C}} \left(\theta - d\tau \int_D d^2\!x\, \sqrt{h} \right)
\ee
where $\wt{\cal C}$ is the projection  of ${\cal C}$ to $\wt{\ca P}$, and $\theta$ is the symplectic potential
on $\wt{\ca P}$ (which is locally
equal to a sum $\sum_i p_i dq^i$
over a complete set of canonically conjugate coordinates).
Thus the reduced
(time-dependent) Hamiltonian is identified as $\wt H(\tau) = \int_D d^2\!x\, \sqrt{h}$, that is, the area of the CMC surface
with 
$K = -\tau$~\cite{york1972role}.

\section{Quantum}

As the reduced phase space does not seem to admit
a natural global coordinate chart, 
the traditional Dirac canonical
quantization rule $\{q, p\} = 1 \mapsto \frac{1}{i\hbar} [\wh q, \wh p] = \wh 1$ cannot be straightforwardly implemented.
Isham developed a generalization of
Dirac's canonical quantization rule that,
rather than being based on 
a preferred coordinate system,
is designed to preserve the structure of a
group of symplectic (canonical) transformations acting 
transitively on the phase space~\cite{isham1984topological, isham1989canonical}. In the 
simple case of  a particle 
on a line $\bb R$, the functions $x$ and $p$ on phase
space, acting as Hamiltonian ``charges", generate the group of phase space translations, which is 
represented projectively, unitarily and irreducibly  in the quantum theory. 
More generally, 
given a group $G$ of symplectic symmetries 
acting on the phase space, we can generate a set of observables whose Poisson algebra closes. These observables are the Hamiltonian charges $Q_i$ associated with the algebra $\fr g$ of $G$, and their Poisson algebra is homomorphic to $\fr g$, up to possible central extensions. If there are central extensions, we extend $G$
to include them as generators, so that the Poisson algebra
is then homomorphic to $\fr g$.
If the group action is transitive then the set $\{Q_i\}$ is complete in the sense that any function on the phase space can be locally written in terms of them. 
Quantization then proceeds by 
replacing the Poisson algebra by a commutator algebra,
$\{Q_i, Q_j\} = c^k_{ij} Q_k \mapsto \frac{1}{i\hbar} [{\wh Q}_i, {\wh Q}_j ]= c^k_{ij} {\wh Q}_k$, and finding unitary irreducible representations of this algebra.

Isham quantization is particularly natural 
when the phase space is the 
cotangent bundle of a homogeneous space, 
$\wt{\ca P}=T^*(K/H)$, where $H$ is a subgroup of
a group $K$. The configuration space 
$K/H$ carries a natural action of $K$ 
that lifts
to the cotangent bundle, and this provides 
``half" of the quantization group. 
There is a simple way to extend this group 
by ``momentum translations'' generated by
charges defined globally on the phase space:
given any function $f$ on $K/H$, the 1-form 
$df$ at every point can be subtracted from 
the momentum 1-forms at that point.
This defines a symplectic map of the 
phase space that is generated by the
function $f$. To define a 
transitive action on the 
phase space together with the $K$-action
one must choose a sufficiently large collection
of such functions; and,
to minimize the inclusion of algebra representations
that fail to produce the desired classical limit, 
this collection of functions should
presumably be as small as possible.
Isham identified
a construction that does exactly this, 
provided $K$ can be linearly represented on a vector space $V$ 
in such a way that at least one of the $K$-orbits  in $V$ is homeomorphic to $K/H$:
linear functions on $V$, i.e.\ elements of the dual $V^*$,
induce on the orbit, and therefore on $K/H$, a suitable collection of functions. 
Together with $K$ the corresponding momentum translations define a transitive group $G = V^* \rtimes K$ of symmetries on $\wt{\ca P}$.\footnote{For the 
example $K/H = SO(3)/SO(2)(=S^2)$,
the $SO(3)$ charges are the components of angular 
momentum, the momentum translations are the Cartesian
coordinates of the $\bb R^3$ in which the 
configuration space $S^2$ is realized as an orbit of $SO(3)$, and the quantizing group is $\bb R^{3*}\rtimes SO(3)$, the Euclidean group in three
dimensions~\cite{e2021particle}.}

In our case, $\wt{\ca P} = T^*\ca Q$, 
where $\ca Q = \diff/\psl$, 
the group $K = \diff$ naturally acts from the left on $\ca Q$,
but we have not found a representation of $\diff$ containing an orbit homeomorphic to $\cal Q$. 
Fortunately, however, for the purpose of identifying a suitable set of functions on 
$\ca Q$
we can take $K$ to be the Virasoro group $\vira$, which is a central extension of $\diff$ and thus can also act on $\ca Q$ (where the central element just acts trivially). The coadjoint representation
of $\vira$, which acts on $V = \dvira$ (where $\avira$ is the Lie algebra of $\vira$), does contain an orbit isomorphic to $\ca Q$~\cite{witten1988coadjoint,lazutkin1975normal,segal1981unitary,alekseev1989path},
hence we can take $G = {(\dvira)}^* \rtimes \vira$ as the group to be quantized. This group is a central extension of BMS${}_3$~\cite{barnich2007classical,oblak2017bms}.\footnote{BMS${}_3$
is familiar as the symmetry of asymptotically Minkowskian spacetime acting on the null cone at future null infinty.
Here it appears as a natural group of symplectic transformations
acting on the phase space of the diamond. 
Perhaps there is a different way to view the reduction of the phase space of the diamond and the action of this group, in terms of the null surfaces that bound the diamond.}

In this way, the quantum theory is based on irreducible unitary (projective) representations of ${(\dvira)}^* \rtimes \vira$. Since this group has the form of a semi-direct product with an abelian factor (here $(\dvira)^*$ with its vector space group structure), we could hope to use Mackey's theory of induced representations to classify the representations~\cite{mackey1969induced}. (Mackey's classification has not been rigorously established
for infinite dimensions, however \cite{mackey1963infinite}.)  Basically, for any $K$-orbit $\ca O$ in ${{(\dvira)}^*}^*$, with 
corresponding
little group $H_{\ca O}$, one can construct a unitary irreducible representation (irrep)
consisting of wavefunctions on $\ca O$ taking values in unitary irreps of $H_{\ca O}$. Note that, modulo issues of infinite-dimensionality, ${{(\dvira)}^*}^* \!\!\!\sim \dvira$, and 
one of the orbits in $\dvira$ is just $\diff/\psl$, so there exist representations given by wavefunctions on $\ca Q$, taking values in unitary irreps of the corresponding little group $\psl \times \bb R$ (where $\bb R$ is the central element of $\vira$). In particular, taking the trivial irrep of $\psl \times \bb R$ gives the usual Hilbert space of  $\bb C$-valued wavefunctions on $\ca Q$,
but it is worth noting that this is only one among a plethora of possibilities. 
Much as the quantization of a relativistic particle
revealed the possibility of intrinsic spin, which is in fact realized in nature,
perhaps
the nontrivial representations of the little
group $\psl \times \bb R$ have 
physical significance for quantum gravity.

We can also think in terms of the representations of the algebra of $G$, 
$\fr g = \avira^c \oplus_S \avira$, where $\avira^c$ is the commutative algebra of momentum translations (which is isomorphic to $(\dvira)^* \sim \avira$ as a vector space).
Note that $\avira = \adiff \oplus_S \bb R$, so its elements can be characterized by a vector field on $S^1$ plus a real number corresponding to the central direction. A convenient basis is defined by Fourier modes of the vector field, that is, 
$L_n = e^{in\theta} \partial_\theta$, with the central element denoted by $R$. Similarly, $\avira^c$ is spanned by elements $A_n = e^{in\theta} \partial_\theta$ and the central element denoted by $T$. 
The algebra reads
\begin{align}
[L_n, L_m] &= i (n - m) L_{n+m} - 4\pi i n^3 \delta_{n+m, 0} R \nonumber\\
[A_n, L_m] &= i (n - m) A_{n+m} - 4\pi i n^3 \delta_{n+m, 0} T \nonumber\\
[A_n, A_m] &= 0 \nonumber\\
[R, ~~\cdot ~~] &= 0 \nonumber\\
[T, ~~\cdot ~~] &= 0 \label{calg}
\end{align}
where $n, m \in \bb Z$.\footnote{Note 
that the Lie algebra bracket for the diffeomophism group is the negative of the Lie bracket of the corresponding vector fields on the manifold.}
We reiterate that the $L$'s and $R$ are associated with the ``configuration translations''
(i.e., the $K$ action), and the $A$'s and $T$ with the ``momentum translations'' (i.e., the $V^*$ action), but note that $R$ and $T$ act trivially on the 
phase space. We find that 
this algebra can be realized by Poisson brackets on the phase space 
with a suitable choice of the charges $P_n$ and $Q_n$ corresponding to $L_n$ and $A_n$, respectively, 
provided that the central charges $R$ and $T$ are realized by the 
constant functions $0$ and $1$, respectively. (This choice of charges is discussed in Sec. \ref{confcoord}.) The resulting Poisson algebra
is
\begin{align}
&\{P_n, P_m\} = i (n - m) P_{n+m} \nonumber\\
&\{Q_n, P_m\} = i (n - m) Q_{n+m} - 4\pi i n^3 \delta_{n+m, 0} \nonumber\\
&\{Q_n, Q_m\} = 0\,. \label{bmsc}
\end{align}
This is a centrally extended
$\fr{bms}_3$ algebra~\cite{barnich2007classical}.
Finally, quantization amounts to associating operators ${\wh P}_n$ and ${\wh Q}_n$ to $P_n$ and $Q_n$, respectively, and replacing $\{~, ~\}$ by $\frac{1}{i\hbar} [~,~]$, 
\begin{align}
&[{\wh P}_n, {\wh P}_m] = \hbar (m - n) {\wh P}_{n+m} \nonumber\\
&[{\wh Q}_n, {\wh P}_m] = \hbar (m - n) {\wh Q}_{n+m} + 4\pi \hbar n^3 \delta_{n+m, 0} \nonumber\\
&[{\wh Q}_n, {\wh Q}_m] = 0\,. \label{bmsq}
\end{align}
The classical charges are not real and instead satisfy $(P_n)^* = P_{-n}$ and $(Q_n)^* = Q_{-n}$, so their associated  operators must satisfy analogous adjoint relations, $({\wh P}_n)^\dag = {\wh P}_{-n}$ and $({\wh Q}_n)^\dag = {\wh Q}_{-n}$.
Some aspects of the 
representation theory of this algebra has been studied recently \cite{oblak2017bms, barnich2014notes,
barnich2015notes,
oblak2015characters, campoleoni2016bms}.

Note that \eqref{bmsq} corresponds to a representation of \eqref{calg} in which 
the quantum Casimir operators $\wh T$ and $\wh R$ 
match the classical values of 1 and 0, respectively.
In the Mackey construction of induced representations of ${(\dvira)}^* \rtimes \vira$
we must therefore select an orbit on which $\wh T$ is represented as the identity and the central $\bb R$ factor in the 
little group is represented trivially.
The natural $\diff/\psl$ orbit is suitable for that purpose~\cite{rodrigo2022},
in which case the wave functions 
transform under a representation
of $\psl$.

\section{Conformal coordinates
and the canonical charges}\label{confcoord}

In this section we briefly introduce the {\it conformal coordinates} which allow us to carry out the reduction process in an explicit fashion, providing the map between the geometrical variables (e.g., spatial metric and extrinsic curvature) and the abstract gauge-invariant variables describing the reduced phase space. Such a map is relevant in understanding the physical/geometrical meaning of observables like the $Q$ and $P$ charges. A treatment including all details is given in \cite{rodrigo2022}. {\it This section is somewhat technical and can be skipped on a first read.}

By virtue of the uniformization theorem, any Riemannian metric $h_{ab}$ on the disc $D$ can be obtained from a reference metric $\bar h_{ab}$ via some conformal transformation. That is, there exists an (orientation-preserving) diffeomorphism $\Psi : D \rightarrow D$ and a positive scalar $\Omega : D \rightarrow \bb R^+$ such that $h_{ab} = \Psi_* \Omega \bar h_{ab}$.
Because of the boundary condition on $h$,  $h|_{\partial D} = \gamma$, the boundary value of $\Omega$ is determined from the boundary action of $\Psi$, $\psi := \Psi|_{\partial D}$, $\left. \Omega \bar h \right|_{\partial D} = \psi^{-1}_* \gamma$. We shall choose the reference disc to be the unit Euclidean disc, so $\bar h = dr^2 + r^2 d\theta^2$ in the usual polar coordinates, and 
choose $\theta$ without loss of generality so as to satisfy
$\gamma = (\ell/2\pi)^2d\theta^2$. Note that, given $h$, $\Psi$ is determined only up to a $\psl$ ambiguity since the transformation can be composed from the right with a conformal isometry of the reference disc, i.e., if $\Phi_*\Theta \bar h = \bar h$ then $(\Psi, \Omega) \circ (\Phi, \Theta) = (\Psi \circ \Phi, \Phi^*\Omega\,\Theta)$ also maps $\bar h$ to $h$.
(We are introducing additional gauge in the description, which is fine since it will be all removed in the end.) We define the ``pull-back'' of $\sigma^{ab}$ 
to the reference disc by $\bar\sigma^{ab} := \Omega^2 \Psi^{-1}_* \sigma^{ab}$, which implies that $\bar\sigma^{ab}$ is
symmetric, traceless and divergenceless 
with respect to $\bar h$ if and only if $\sigma^{ab}$ has the same properties 
with respect to $h$. So far we have a ``change of coordinates'' from $(h_{ab}, \sigma^{ab})$ 
to $(\Psi, \Omega, \bar \sigma^{ab})$.  
Imposing the scalar constraint leads to a Lichnerowicz equation for $\Omega$, and
the boundary value of $\Omega$ is determined from $\psi$ (and $\gamma$); since that equation has a unique solution for $\Omega$,
given $\psi$ and $\bar\sigma^{ab}$, the constraint surface in phase space
can be parametrized by $(\Psi, \bar\sigma^{ab})$, where $\bar\sigma^{ab}$ is symmetric, traceless and divergenceless with respect to $\bar h$. This space of $\bar \sigma$'s is isomorphic to a subspace of dual vector fields $\hat\sigma$ on the boundary $S^1$;
given the form of the symplectic
structure, it is natural to realize 
the isomorphism as $\hat\sigma(\xi) := \int\!d\theta\, \hat\sigma(\theta) \xi(\theta) := -2 \int\!d\theta\, \bar\sigma^{ab} n_a \xi_b$, where $\xi = \xi(\theta)\, \partial_\theta$ is a vector field on the boundary and $n$ is the unit outward-pointing normal vector field on the boundary.
In this realization of the isomorphism, the space of $\hat\sigma$'s
is missing the Fourier modes $1$, $\sin\theta$, $\cos\theta$, since they annihilate the vector fields $\xi(\theta) = 1,\, \sin\theta,\, \cos\theta$.
Via this isomorphism, the constraint surface can be parametrized by $(\Psi, \hat\sigma)$. It is clear from the pre-symplectic form that any two $\Psi$'s with the same boundary action $\psi$ are gauge-equivalent, 
so we can quotient out the bulk diffeomorphisms and obtain a partially-reduced phase space coordinatized by $(\psi, \hat\sigma)$. By further inspection of the symplectic form one discovers that there remains
a $\psl$ group of gauge transformations, which acts on $\psi$ from the right and on $\hat\sigma$ via the coadjoint action. 
The quotient under this group finally leads to the reduced phase space $T^*[\diff/\psl]$. 

The canonical charges can be explicitly expressed in terms of the $(\psi, \hat\sigma)$ variables. (Only the results are presented here; the derivation can be found in \cite{rodrigo2022}.)
As the canonical group acts on the phase space, each element $\zeta$ of the Lie algebra induces a vector field $X_\zeta$ on the phase space; the
a corresponding Hamiltonian charge $H_\zeta$ is a
solution of $\delta H_\zeta = - i_{X_\zeta}\omega$, where $\delta$ denotes the exterior derivative on phase space and $i_X \omega$ is the insertion of $X$ into the first slot of the symplectic form $\omega$.
The ``momentum'' ($P$) charges are associated with the $\vira$ part of the group, acting as configuration space ``translations'', therefore corresponding to 
 algebra elements purely in the $\avira$ factor of $\fr g = \avira^c \oplus_S \avira$. If $\wh\xi = (\xi(\theta)\partial_\theta, \xi_0) \in \avira$, where $\xi_0$ is the central component, 
then
\be\label{Pexpr}
P_{\wh\xi}(\psi, \hat\sigma) = \int\!d\t\, \frac{\hat\sigma(\t)}{\psi'(\t)} \xi(\psi(\t))
\ee
In the earlier notation, $P_n := P_{\wh\xi = (e^{in\theta} \partial_\theta, 0)}$, and the central charge $R := P_{\wh\xi = (0, 1)} = 0$.
The ``position'' ($Q$) charges are associated with the ${(\dvira)}^*$ part of the group, acting as ``vertical translations'' on phase space, thus corresponding to algebra elements purely in the $\avira^c$ factor of $\fr{g}$.
If $\wh\eta = (\eta(\theta)\partial_\theta, \eta_0) \in \avira^c$, where $\eta_0$ is the central component, then
\be\label{Q}
Q_{\wh\eta}(\psi, \sigma) = \int\!d\t\, \frac{1 - 2 S[\psi](\t)}{\psi'(\t)} \eta(\psi(\t)) + \eta_0
\ee
where $S[\psi](\t) := \psi'''(\t)/\psi'(\t) - \frac{3}{2} ( \psi''(\t)/\psi'(\t))^2$ is the Schwarzian derivative of $\psi$.
In the earlier notation, $Q_n := Q_{\wh\eta = (e^{in\theta} \partial_\theta, 0)}$; and the central charge $T := Q_{\wh\eta = (0,1)} = 1$.

It is straightforward to express the $P_\xi$ charges in terms of the physical spatial metric and extrinsic curvature.
This can be done by direct manipulation of expression \eqref{Pexpr}, basically by reversing the map from the reference disc
variables $(\bar h_{ab}, \bar \sigma^{ab})$
to the physical disc
variables $(h_{ab}, \sigma^{ab})$
so as to express
$(\psi, \hat\sigma)$ in terms of $(h_{ab}, \sigma^{ab})$. Instead of going through this formal derivation (which can be found in \cite{rodrigo2022}), we can infer the answer by noticing that the charge must descend from a function on the unreduced phase space that generates a corresponding diffeomorphism on the spatial slice. We know that this charge must be related to $\int\!d^2x\, \pi^{ab} \pounds_\xi h_{ab}$, where $\xi$ is now an arbitrary extension of the boundary vector field to the disc.  However this function alone generates a pure diffeomorphism on the ADM phase space and thus does not generally respect the boundary conditions on the induced metric (unless $\xi$ is an isometry of the boundary metric). That can be fixed by adding a constraint term which generates a compensating Weyl transformation.
The appropriate constraint here comes from the gauge-fixing of time $\tau = -K$, that is, $P_\xi = - \int\!d^2x\, \pi^{ab} \pounds_\xi h_{ab} + \int\!d^2x \sqrt{h}\, \zeta (K + \tau)$ for some scalar $\zeta$. 
When this expression is evaluated imposing 
the CMC gauge condition
and the vector constraint $\nabla_a\pi^{ab}=0$, it reduces to 
$P_\xi = -2 \int\!d^2x \sqrt{h}\, \sigma^{ab} \nabla_a \xi_b$.
This inferred form can be shown to agree with the pull-back 
to the (constrained, gauge-fixed) ADM phase space 
of the $P_\xi$'s defined in \eqref{Pexpr}.
Using Stoke's theorem we get $P_{\xi} =  - 2\int_\partial\!ds\,  K_{ab}  n^a \xi^b$, where $n$ is the unit outward-pointing normal vector field at the boundary of the disc,
and $ds$ is the proper length along the boundary.
Restoring the factor of $16\pi G$ that had previously been set to unity, this becomes $P_{\xi} =  -\frac{1}{8\pi G} \int_\partial\!ds\, K_{ab}  n^a \xi^b$. 
The vector field $\xi$ that labels the charge $P_0$ is
$\partial_\theta$ on the reference disk.
In terms of the
vector field $t^a$ tangent to the boundary, with unit norm with respect to the physical metric $\gamma$, we have on the boundary $\partial_\theta=
\frac{\ell}{2\pi}t^a$,
hence $P_0 =  - \frac{\ell}{16\pi^2 \ell_P} \int_\partial\!ds\,  K_{ab}  n^a t^b$. If $u$ is the unit future-pointing vector field normal to the CMC slice, then $P_0 =  - \frac{\ell}{16\pi^2 \ell_P} \int_\partial\!ds\,  \boldsymbol\nabla_b u_a  n^a t^b$. Integrating by parts we conclude that $P_0 = \frac{\ell}{16\pi^2 \ell_P} \int_\partial\!ds\, u_a t^b  \boldsymbol\nabla_b n^a = \frac{\ell}{16\pi^2 \ell_P} \ca T$, where $\ca T$ is the twist of the boundary loop, as embedded in the spacetime, which is defined as the integral of the torsion $u_a t^b  \boldsymbol\nabla_b n^a$ with respect to proper length.

Regarding the appearance of the 
Schwarzian in the expression \eqref{Q}
for the $Q_\eta$ charges we offer here a brief explanation. When the configuration space is embedded as a coadjoint orbit in $\dvira$, each point $x \in \ca Q$ corresponds to an element of $\dvira$. In this context, the charge $Q_\eta$ evaluated at $x$ is the value of the dual vector $x \in \dvira$ acting on the vector $\eta \in \avira$, i.e., $Q_\eta(x) = x(\eta)$. The point $x$ is 
labeled by a diffeomorphism $\psi$, relative to 
a reference point $x_0 \in \ca Q$, via the coadjoint action
$x = \text{\sl coad}_\psi x_0$. (Of course this labelling system is not one-to-one because $x_0$ is invariant under a $\psl$ subgroup of $\diff$.) 
 This yields the expression $Q_\eta(x) = \text{\sl coad}_\psi x_0(\eta)$, which for a simple choice of $x_0$
 corresponds to \eqref{Q}. The Schwarzian 
appears in this expression because it figures in 
the coadjoint action.

Note that the $Q$'s do not depend on $\hat\sigma$ and, as can be shown from basic properties of the Schwarzian derivative, depend only on the right $\psl$ equivalence classes $[\psi] \in \diff/\psl$. A given spatial metric $h$ uniquely determines one such equivalence classes $[\psi]$, and one class $[\psi]$ determines a spatial metric up to boundary-trivial conformal transformations, $[h] = [\Phi_*\Theta h]$, where $\Phi \in \text{Diff}^+(D)$ acts as the identity on the boundary and the function $\Theta$ is $1$ at the boundary. Therefore, the $Q$ charges evidently 
depend only on the conformal class
of the spatial metric. 

It can be shown that $Q_0$ is bounded from above, attaining a maximum value of $2\pi$ when $[\psi] = [I]$. In that configuration, $Q_{\wh\eta} = \int\!d\theta\, \eta(\theta) + \eta_0$, hence all $Q_n$ with $n\ne 0$ vanish. Classically it corresponds to a spatial geometry that is related to the round disc by a boundary-trivial conformal transformation.

\section{Spin/Twist}

An interesting observable to discuss in more detail is $P_0$. It is the ``zero Fourier mode'' of $\diff \subset \vira$, i.e., it generates the $SO(2)$ subgroup of rotations, suggesting that it corresponds to the {\sl spin} of the diamond. This interpretation can be further strengthened by noticing that it is precisely (minus) the on-shell value of the ADM charge associated with a vanishing lapse and a shift that acts as an isometry of the boundary loop. 
The charge $P_0$ generates not only a symmetry of the symplectic form (as do all of the $P$'s and $Q$'s), but also a true dynamical symmetry. That is, it commutes with the CMC time evolution Hamiltonian
(defined below \eqref{action}), $[P_0, \wt H] = 0$, as will become clear presently.
We have argued that the physical states correspond (classically) to shapes of diamonds embedded in $\ads$ (or Mink${}_3$ if $\Lambda=0$), with boundary length $\ell$, so $P_0$ must correspond to some aspect of the shape. As shown in Sec. \ref{confcoord},
it turns out that $P_0$ is proportional to the {\sl twist} $\ca T$ of the diamond boundary loop,
i.e.\  the loop integral
(with respect to proper length) of the 
torsion of the curve
(as embedded in the spacetime).
The twist can also be interpreted as  
the holonomy of Fermi-Walker transport
of an orthogonal frame around the loop, i.e., 
the (hyperbolic) angle of the boost relating the
final frame to the initial one. 
The precise relation
(which is obtained using the
previously mentioned ``conformal coordinates" 
characterization of the reduced phase space)
is
\be\label{P0}
P_0 = \frac{\ell}{16\pi^2G} \ca T
\ee
Note that the twist of the boundary
is clearly independent of the CMC slice
of the diamond, hence it is time independent and thus commutes with $\wt H$ as stated above.\footnote{This relationship between twist and spin seems to be related to a result in \cite{donnelly2021gravitational}. Working in an extended phase space including edge modes in 3+1 spacetime dimensions, they find that the generator of volume-preserving diffeomorphisms of the ``corner'', $S^2$, is essentially the curvature of the natural connection on the normal bundle of $S^2$ (as embedded in the ambient spacetime). In our case the corner 
is the boundary loop, $S^1$; volume-preserving diffeomorphisms are just the isometries of the boundary metric; and, although the curvature of the normal bundle connection vanishes (because $S^1$ is 1-dimensional), there is a non-trivial holonomy (around the loop) which is equal to the twist.}

At the quantum level, note that the Poisson brackets \eqref{bmsc} imply $[{\wh P}_0, {\wh P}_n] = n\hbar{} {\wh P}_n$ and $[{\wh P}_0, {\wh Q}_n] = n\hbar {\wh Q}_n$, so the $P$'s and $Q$'s act as ladder operators for ${\wh P}_0$. That is, if $|s\ra$ is an eigenvector of ${\wh P}_0$ with eigenvalue $s\hbar$, then ${\wh P}_n|s\ra$ and ${\wh Q}_n|s\ra$ have eigenvalue $(s+n)\hbar$. Since the $P$'s and $Q$'s are represented irreducibly in the Hilbert space, the spectrum of ${\wh P}_0$ is  $\{(s+n)\hbar, \forall\, n\in \bb Z\}$,
where
without loss of generality we can take $s \in [0, 1)$.
Classically, $\tau$-time reversal flips  the sign of $P_0$; if this (anti-symplectic) symmetry of the phase space is represented by an anti-unitary transformation in the quantum theory---as one might
expect given that the classical Hamiltonian is invariant under this symmetry---then in particular the spectrum of ${\wh P}_0$ will be symmetric under sign reversal. In this case, only $s=0$ and $s=\frac12$ are allowed. 
From formula \eqref{P0} we conclude that the twist is quantized as
\be
\ca T = \frac{16\pi^2 \ell_P}{\ell} (s + n)\,,\quad n\in \bb Z
\ee
where (in 3d) $\ell_P = \hbar G$ is the Planck length, in units with $c=1$.
In the classical limit 
$\ell \gg \ell_P$, the twist quantum is very small, so that a continuum
of twist values is recovered.

\section{Discussion}

We studied quantization of 
causal diamonds of fixed boundary length in
pure 2+1 dimensional general relativity
gravity with a nonpositive 
cosmological constant,
via the reduced phase space approach.
The low dimensionality allowed us to solve the constraints exactly and remove all the gauge ambiguities, resulting in the phase space $T^*\ca Q$ with $\ca Q=\diff/\psl$.
Further, this phase space could be quantized exactly, at the kinematical level, with all the rigor and generality of Isham's group quantization scheme. We ended up with a classification of all possible quantizations based on irreducible unitary representations of the $\fr{bms}_3$ algebra.
This differs from the canonical quantization of pure asymptotically $\ads$ (with trivial topology) based on the group $\vira \times \vira$ \cite{maloney2010quantum}, whose algebra is $\avira \oplus \avira$.
Note that although the quantization groups differ the phase spaces are the same, since $T^*\ca Q \sim \ca Q\times\ca Q$~\cite{Edward,Scarinci:2011np}. 

The quantization was strictly kinematical because only the canonical charges
$Q_n$ and $P_n$
(``coordinates'' on phase space) have been quantized. This was sufficient to reveal that the spin
of the diamond, or equivalently the
twist of the diamond boundary loop, is quantized in integer or half-integer multiples of
$16\pi^2 \ell_P/\ell$. 
To fully characterize the 
quantum theory  one must also
represent the Hamiltonian $\wt H$
that generates evolution in CMC time. This Hamiltonian is, however, a very complicated function on the reduced phase space, for which we have not found any preferred operator ordering or even any explicit expression
in terms of the canonical charges.
It may be that progress could be made 
using a perturbative approach.
There are certain regimes where the Hamiltonian simplifies, even as much as becoming 
``free" (quadratic in $P_n$) 
in the limit $\ell\gg |\Lambda|^{-1/2}$ when 
the maximal slice is nearly
a hyperbolic disc.
This includes the case where the boundary loop approaches the boundary of AdS, in which the diamond approaches a ``Wheeler-DeWitt patch" of AdS.
It would be interesting to explore such regimes, and in particular the possible connection to quantization of the diamond from the perspective of AdS/CFT duality (and its TTbar deformations).

Another important open question is
the geometrical meaning of the 
charges $Q_n$. Unlike the $P_n$,
which have a simple interpretation
as Fourier components of the torsion of the boundary curve, the $Q_n$ are related
to the shape of the diamond in a complicated, implicit fashion.
As explained in Sec. \ref{confcoord} we know that
the $Q$ charges depend only on the configuration space variables $[\psi] \in \diff/\psl$, which implies that they depend only on the conformal class of the spacial metric, $[h_{ab}]$, where two metrics are identified if they can be related by a conformal transformation that is trivial on the boundary. But despite 
some effort, we have not yet been able to express $Q_n$ directly in terms of the spatial conformal metric.
In the asymptotically flat case, whose group of symmetries at null infinity is also BMS${}_3$, $-Q_0$ plays the role of energy (i.e., the generator of $u$-coordinate translations, up to a scaling factor), so by analogy this suggests that $- Q_0$ 
should be some sort of quasilocal mass. In fact, it is noteworthy that for many representations of ${(\dvira)}^* \rtimes \vira$, including the one associated with the orbit $\diff/\psl$ with $T=1$, $-Q_0$ is bounded from below and unbounded from above. In the case of the $\diff/\psl$ orbit the minimum value of $-Q_0$ is equal to $-2\pi$, and it is attained by a (non-normalizable) state 
corresponding to a wavefunction localized at $[\psi]=[I]$, i.e., 
at the spatial geometry conformal to a flat round disc.  
 
One would also 
like to understand what is 
the nature of a ``quantum causal diamond'', 
given that the
classical ``spacetime shape" interpretation, which 
requires that $Q_n$ and $P_n$ are all
simultaneously specified, fails to make sense
in the quantum theory. 
(We note that there are certain observables that do commute among themselves, such as the set including $P_0$, $Q_0$ and any operators of the form $Q_{n_1} Q_{n_2} \cdots$ such that $n_1 + n_2 + \cdots = 0$; some of these operators are actually self-adjoint, like $Q_{-n}Q_n$ for all $n$.)
A perhaps related question is
whether the quantized theory depends
upon the CMC time gauge choice used
for the phase space reduction.
Finally, it might be interesting to analyse the system using the 
formulation of this gravity theory as a pair of $SL(2,R)$ Chern-Simons theories~\cite{Achucarro:1986uwr,witten19882+}.
The fixed metric boundary condition that we have imposed would be a complicated condition that couples those two theories.

\begin{acknowledgments}
We are grateful to Stefano Antonini, 
Luis Apolo, Abhay Ashtekar, Batoul Banihashemi,
Steve Carlip, Gong Cheng, 
Marc Henneaux, Jim Isenberg, 
Alex Maloney, 
Blagoje Oblak,
Pranav Pulakkat, Gabor Sarosi, Antony Speranza,
Yixu Wang and Edward Witten for helpful discussions.
This research was supported in part by the National Science Foundation under Grants PHY-1708139 and 
PHY-2012139 at UMD 
and PHY-1748958 at KITP.
\end{acknowledgments}

\bibliography{Diamond}

\begin{thebibliography}{50}%
\makeatletter
\providecommand \@ifxundefined [1]{%
 \@ifx{#1\undefined}
}%
\providecommand \@ifnum [1]{%
 \ifnum #1\expandafter \@firstoftwo
 \else \expandafter \@secondoftwo
 \fi
}%
\providecommand \@ifx [1]{%
 \ifx #1\expandafter \@firstoftwo
 \else \expandafter \@secondoftwo
 \fi
}%
\providecommand \natexlab [1]{#1}%
\providecommand \enquote  [1]{``#1''}%
\providecommand \bibnamefont  [1]{#1}%
\providecommand \bibfnamefont [1]{#1}%
\providecommand \citenamefont [1]{#1}%
\providecommand \href@noop [0]{\@secondoftwo}%
\providecommand \href [0]{\begingroup \@sanitize@url \@href}%
\providecommand \@href[1]{\@@startlink{#1}\@@href}%
\providecommand \@@href[1]{\endgroup#1\@@endlink}%
\providecommand \@sanitize@url [0]{\catcode `\\12\catcode `\$12\catcode
  `\&12\catcode `\#12\catcode `\^12\catcode `\_12\catcode `\%12\relax}%
\providecommand \@@startlink[1]{}%
\providecommand \@@endlink[0]{}%
\providecommand \url  [0]{\begingroup\@sanitize@url \@url }%
\providecommand \@url [1]{\endgroup\@href {#1}{\urlprefix }}%
\providecommand \urlprefix  [0]{URL }%
\providecommand \Eprint [0]{\href }%
\providecommand \doibase [0]{https://doi.org/}%
\providecommand \selectlanguage [0]{\@gobble}%
\providecommand \bibinfo  [0]{\@secondoftwo}%
\providecommand \bibfield  [0]{\@secondoftwo}%
\providecommand \translation [1]{[#1]}%
\providecommand \BibitemOpen [0]{}%
\providecommand \bibitemStop [0]{}%
\providecommand \bibitemNoStop [0]{.\EOS\space}%
\providecommand \EOS [0]{\spacefactor3000\relax}%
\providecommand \BibitemShut  [1]{\csname bibitem#1\endcsname}%
\let\auto@bib@innerbib\@empty
%</preamble>
\bibitem [{\citenamefont {Isham}(1993)}]{isham1993canonical}%
  \BibitemOpen
  \bibfield  {author} {\bibinfo {author} {\bibfnamefont {C.~J.}\ \bibnamefont
  {Isham}},\ }\bibfield  {title} {\bibinfo {title} {Canonical quantum gravity
  and the problem of time},\ }in\ \href@noop {} {\emph {\bibinfo {booktitle}
  {Integrable systems, quantum groups, and quantum field theories}}}\ (\bibinfo
   {publisher} {Springer},\ \bibinfo {year} {1993})\ pp.\ \bibinfo {pages}
  {157--287}\BibitemShut {NoStop}%
\bibitem [{\citenamefont {Kucha{\v{r}}}(2011)}]{kuchavr2011time}%
  \BibitemOpen
  \bibfield  {author} {\bibinfo {author} {\bibfnamefont {K.~V.}\ \bibnamefont
  {Kucha{\v{r}}}},\ }\bibfield  {title} {\bibinfo {title} {Time and
  interpretations of quantum gravity},\ }\href@noop {} {\bibfield  {journal}
  {\bibinfo  {journal} {International Journal of Modern Physics D}\ }\textbf
  {\bibinfo {volume} {20}},\ \bibinfo {pages} {3} (\bibinfo {year}
  {2011})}\BibitemShut {NoStop}%
\bibitem [{\citenamefont {Anderson}(2012)}]{anderson2012problem}%
  \BibitemOpen
  \bibfield  {author} {\bibinfo {author} {\bibfnamefont {E.}~\bibnamefont
  {Anderson}},\ }\bibfield  {title} {\bibinfo {title} {Problem of time in
  quantum gravity},\ }\href@noop {} {\bibfield  {journal} {\bibinfo  {journal}
  {Annalen der Physik}\ }\textbf {\bibinfo {volume} {524}},\ \bibinfo {pages}
  {757} (\bibinfo {year} {2012})}\BibitemShut {NoStop}%
\bibitem [{\citenamefont {Witten}(1988{\natexlab{a}})}]{witten19882+}%
  \BibitemOpen
  \bibfield  {author} {\bibinfo {author} {\bibfnamefont {E.}~\bibnamefont
  {Witten}},\ }\bibfield  {title} {\bibinfo {title} {2+1 dimensional gravity as
  an exactly soluble system},\ }\href@noop {} {\bibfield  {journal} {\bibinfo
  {journal} {Nuclear Physics B}\ }\textbf {\bibinfo {volume} {311}},\ \bibinfo
  {pages} {46} (\bibinfo {year} {1988}{\natexlab{a}})}\BibitemShut {NoStop}%
\bibitem [{\citenamefont {Witten}(1989)}]{witten1989topology}%
  \BibitemOpen
  \bibfield  {author} {\bibinfo {author} {\bibfnamefont {E.}~\bibnamefont
  {Witten}},\ }\bibfield  {title} {\bibinfo {title} {Topology-changing
  amplitudes in 2+1 dimensional gravity},\ }\href@noop {} {\bibfield  {journal}
  {\bibinfo  {journal} {Nuclear Physics B}\ }\textbf {\bibinfo {volume}
  {323}},\ \bibinfo {pages} {113} (\bibinfo {year} {1989})}\BibitemShut
  {NoStop}%
\bibitem [{\citenamefont {Moncrief}(1989)}]{moncrief1989reduction}%
  \BibitemOpen
  \bibfield  {author} {\bibinfo {author} {\bibfnamefont {V.}~\bibnamefont
  {Moncrief}},\ }\bibfield  {title} {\bibinfo {title} {{Reduction of the
  Einstein equations in 2+1 dimensions to a Hamiltonian system over
  Teichm{\"u}ller space}},\ }\href@noop {} {\bibfield  {journal} {\bibinfo
  {journal} {Journal of Mathematical Physics}\ }\textbf {\bibinfo {volume}
  {30}},\ \bibinfo {pages} {2907} (\bibinfo {year} {1989})}\BibitemShut
  {NoStop}%
\bibitem [{\citenamefont {Moncrief}(1990)}]{moncrief1990solvable}%
  \BibitemOpen
  \bibfield  {author} {\bibinfo {author} {\bibfnamefont {V.}~\bibnamefont
  {Moncrief}},\ }\bibfield  {title} {\bibinfo {title} {{How solvable is
  (2+1)-dimensional Einstein gravity?}},\ }\href@noop {} {\bibfield  {journal}
  {\bibinfo  {journal} {Journal of Mathematical Physics}\ }\textbf {\bibinfo
  {volume} {31}},\ \bibinfo {pages} {2978} (\bibinfo {year}
  {1990})}\BibitemShut {NoStop}%
\bibitem [{\citenamefont {Fischer}\ and\ \citenamefont
  {Moncrief}(1997)}]{fischer1997hamiltonian}%
  \BibitemOpen
  \bibfield  {author} {\bibinfo {author} {\bibfnamefont {A.~E.}\ \bibnamefont
  {Fischer}}\ and\ \bibinfo {author} {\bibfnamefont {V.}~\bibnamefont
  {Moncrief}},\ }\bibfield  {title} {\bibinfo {title} {{Hamiltonian reduction
  of Einstein's equations of general relativity}},\ }\href@noop {} {\bibfield
  {journal} {\bibinfo  {journal} {Nuclear Physics B-Proceedings Supplements}\
  }\textbf {\bibinfo {volume} {57}},\ \bibinfo {pages} {142} (\bibinfo {year}
  {1997})}\BibitemShut {NoStop}%
\bibitem [{\citenamefont {Ashtekar}\ \emph {et~al.}(1989)\citenamefont
  {Ashtekar}, \citenamefont {Husain}, \citenamefont {Rovelli}, \citenamefont
  {Samuel},\ and\ \citenamefont {Smolin}}]{ashtekar19892+}%
  \BibitemOpen
  \bibfield  {author} {\bibinfo {author} {\bibfnamefont {A.}~\bibnamefont
  {Ashtekar}}, \bibinfo {author} {\bibfnamefont {V.}~\bibnamefont {Husain}},
  \bibinfo {author} {\bibfnamefont {C.}~\bibnamefont {Rovelli}}, \bibinfo
  {author} {\bibfnamefont {J.}~\bibnamefont {Samuel}},\ and\ \bibinfo {author}
  {\bibfnamefont {L.}~\bibnamefont {Smolin}},\ }\bibfield  {title} {\bibinfo
  {title} {2+1 quantum gravity as a toy model for the 3+1 theory},\ }\href@noop
  {} {\bibfield  {journal} {\bibinfo  {journal} {Classical and Quantum
  Gravity}\ }\textbf {\bibinfo {volume} {6}},\ \bibinfo {pages} {L185}
  (\bibinfo {year} {1989})}\BibitemShut {NoStop}%
\bibitem [{\citenamefont {Hosoya}\ and\ \citenamefont
  {Nakao}(1990)}]{hosoya19902+}%
  \BibitemOpen
  \bibfield  {author} {\bibinfo {author} {\bibfnamefont {A.}~\bibnamefont
  {Hosoya}}\ and\ \bibinfo {author} {\bibfnamefont {K.}~\bibnamefont {Nakao}},\
  }\bibfield  {title} {\bibinfo {title} {(2+1)-dimensional pure gravity for an
  arbitrary closed initial surface},\ }\href@noop {} {\bibfield  {journal}
  {\bibinfo  {journal} {Classical and Quantum Gravity}\ }\textbf {\bibinfo
  {volume} {7}},\ \bibinfo {pages} {163} (\bibinfo {year} {1990})}\BibitemShut
  {NoStop}%
\bibitem [{\citenamefont {Carlip}(2003)}]{Carlip:1998uc}%
  \BibitemOpen
  \bibfield  {author} {\bibinfo {author} {\bibfnamefont {S.}~\bibnamefont
  {Carlip}},\ }\href@noop {} {\emph {\bibinfo {title} {{Quantum gravity in 2+1
  dimensions}}}},\ Cambridge Monographs on Mathematical Physics\ (\bibinfo
  {publisher} {Cambridge University Press},\ \bibinfo {year}
  {2003})\BibitemShut {NoStop}%
\bibitem [{\citenamefont {Carlip}(2005{\natexlab{a}})}]{carlip2005quantum}%
  \BibitemOpen
  \bibfield  {author} {\bibinfo {author} {\bibfnamefont {S.}~\bibnamefont
  {Carlip}},\ }\bibfield  {title} {\bibinfo {title} {Quantum gravity in 2+1
  dimensions: the case of a closed universe},\ }\href@noop {} {\bibfield
  {journal} {\bibinfo  {journal} {Living Reviews in Relativity}\ }\textbf
  {\bibinfo {volume} {8}},\ \bibinfo {pages} {1} (\bibinfo {year}
  {2005}{\natexlab{a}})}\BibitemShut {NoStop}%
\bibitem [{\citenamefont {Kraus}\ \emph {et~al.}(2021)\citenamefont {Kraus},
  \citenamefont {Monten},\ and\ \citenamefont {Myers}}]{kraus20213d}%
  \BibitemOpen
  \bibfield  {author} {\bibinfo {author} {\bibfnamefont {P.}~\bibnamefont
  {Kraus}}, \bibinfo {author} {\bibfnamefont {R.}~\bibnamefont {Monten}},\ and\
  \bibinfo {author} {\bibfnamefont {R.~M.}\ \bibnamefont {Myers}},\ }\bibfield
  {title} {\bibinfo {title} {3d gravity in a box},\ }\href@noop {} {\bibfield
  {journal} {\bibinfo  {journal} {SciPost Physics}\ }\textbf {\bibinfo {volume}
  {11}},\ \bibinfo {pages} {070} (\bibinfo {year} {2021})}\BibitemShut
  {NoStop}%
\bibitem [{\citenamefont {Adami}\ \emph {et~al.}(2020)\citenamefont {Adami},
  \citenamefont {Hosseinzadeh},\ and\ \citenamefont
  {Sheikh-Jabbari}}]{adami2020sliding}%
  \BibitemOpen
  \bibfield  {author} {\bibinfo {author} {\bibfnamefont {H.}~\bibnamefont
  {Adami}}, \bibinfo {author} {\bibfnamefont {V.}~\bibnamefont
  {Hosseinzadeh}},\ and\ \bibinfo {author} {\bibfnamefont {M.}~\bibnamefont
  {Sheikh-Jabbari}},\ }\bibfield  {title} {\bibinfo {title} {Sliding surface
  charges on {AdS3}},\ }\href@noop {} {\bibfield  {journal} {\bibinfo
  {journal} {Physics Letters B}\ }\textbf {\bibinfo {volume} {806}},\ \bibinfo
  {pages} {135503} (\bibinfo {year} {2020})}\BibitemShut {NoStop}%
\bibitem [{\citenamefont {Ebert}\ \emph {et~al.}(2022)\citenamefont {Ebert},
  \citenamefont {Hijano}, \citenamefont {Kraus}, \citenamefont {Monten},\ and\
  \citenamefont {Myers}}]{ebert2022field}%
  \BibitemOpen
  \bibfield  {author} {\bibinfo {author} {\bibfnamefont {S.}~\bibnamefont
  {Ebert}}, \bibinfo {author} {\bibfnamefont {E.}~\bibnamefont {Hijano}},
  \bibinfo {author} {\bibfnamefont {P.}~\bibnamefont {Kraus}}, \bibinfo
  {author} {\bibfnamefont {R.}~\bibnamefont {Monten}},\ and\ \bibinfo {author}
  {\bibfnamefont {R.~M.}\ \bibnamefont {Myers}},\ }\bibfield  {title} {\bibinfo
  {title} {Field theory of interacting boundary gravitons},\ }\href@noop {}
  {\bibfield  {journal} {\bibinfo  {journal} {arXiv preprint arXiv:2201.01780}\
  } (\bibinfo {year} {2022})}\BibitemShut {NoStop}%
\bibitem [{\citenamefont {Brown}\ and\ \citenamefont
  {Henneaux}(1986)}]{brown1986central}%
  \BibitemOpen
  \bibfield  {author} {\bibinfo {author} {\bibfnamefont {J.~D.}\ \bibnamefont
  {Brown}}\ and\ \bibinfo {author} {\bibfnamefont {M.}~\bibnamefont
  {Henneaux}},\ }\bibfield  {title} {\bibinfo {title} {Central charges in the
  canonical realization of asymptotic symmetries: an example from three
  dimensional gravity},\ }\href@noop {} {\bibfield  {journal} {\bibinfo
  {journal} {Communications in Mathematical Physics}\ }\textbf {\bibinfo
  {volume} {104}},\ \bibinfo {pages} {207} (\bibinfo {year}
  {1986})}\BibitemShut {NoStop}%
\bibitem [{\citenamefont {Freidel}\ \emph {et~al.}(2004)\citenamefont
  {Freidel}, \citenamefont {Kowalski-Glikman},\ and\ \citenamefont
  {Smolin}}]{freidel20042+}%
  \BibitemOpen
  \bibfield  {author} {\bibinfo {author} {\bibfnamefont {L.}~\bibnamefont
  {Freidel}}, \bibinfo {author} {\bibfnamefont {J.}~\bibnamefont
  {Kowalski-Glikman}},\ and\ \bibinfo {author} {\bibfnamefont {L.}~\bibnamefont
  {Smolin}},\ }\bibfield  {title} {\bibinfo {title} {2+1 gravity and doubly
  special relativity},\ }\href@noop {} {\bibfield  {journal} {\bibinfo
  {journal} {Physical Review D}\ }\textbf {\bibinfo {volume} {69}},\ \bibinfo
  {pages} {044001} (\bibinfo {year} {2004})}\BibitemShut {NoStop}%
\bibitem [{\citenamefont {Carlip}(2005{\natexlab{b}})}]{carlip2005conformal}%
  \BibitemOpen
  \bibfield  {author} {\bibinfo {author} {\bibfnamefont {S.}~\bibnamefont
  {Carlip}},\ }\bibfield  {title} {\bibinfo {title} {Conformal field theory,
  (2+1)-dimensional gravity and the {BTZ} black hole},\ }\href@noop {}
  {\bibfield  {journal} {\bibinfo  {journal} {Classical and Quantum Gravity}\
  }\textbf {\bibinfo {volume} {22}},\ \bibinfo {pages} {R85} (\bibinfo {year}
  {2005}{\natexlab{b}})}\BibitemShut {NoStop}%
\bibitem [{\citenamefont {Witten}(2007)}]{witten2007three}%
  \BibitemOpen
  \bibfield  {author} {\bibinfo {author} {\bibfnamefont {E.}~\bibnamefont
  {Witten}},\ }\bibfield  {title} {\bibinfo {title} {Three-dimensional gravity
  revisited},\ }\href@noop {} {\bibfield  {journal} {\bibinfo  {journal} {arXiv
  preprint arXiv:0706.3359}\ } (\bibinfo {year} {2007})}\BibitemShut {NoStop}%
\bibitem [{\citenamefont {Maloney}\ and\ \citenamefont
  {Witten}(2010)}]{maloney2010quantum}%
  \BibitemOpen
  \bibfield  {author} {\bibinfo {author} {\bibfnamefont {A.}~\bibnamefont
  {Maloney}}\ and\ \bibinfo {author} {\bibfnamefont {E.}~\bibnamefont
  {Witten}},\ }\bibfield  {title} {\bibinfo {title} {Quantum gravity partition
  functions in three dimensions},\ }\href@noop {} {\bibfield  {journal}
  {\bibinfo  {journal} {Journal of High Energy Physics}\ }\textbf {\bibinfo
  {volume} {2010}},\ \bibinfo {pages} {1} (\bibinfo {year} {2010})}\BibitemShut
  {NoStop}%
\bibitem [{\citenamefont {Scarinci}\ and\ \citenamefont
  {Krasnov}(2013)}]{Scarinci:2011np}%
  \BibitemOpen
  \bibfield  {author} {\bibinfo {author} {\bibfnamefont {C.}~\bibnamefont
  {Scarinci}}\ and\ \bibinfo {author} {\bibfnamefont {K.}~\bibnamefont
  {Krasnov}},\ }\bibfield  {title} {\bibinfo {title} {{The universal phase
  space of $AdS_3$ gravity}},\ }\href@noop {} {\bibfield  {journal} {\bibinfo
  {journal} {Commun. Math. Phys.}\ }\textbf {\bibinfo {volume} {322}},\
  \bibinfo {pages} {167} (\bibinfo {year} {2013})}\BibitemShut {NoStop}%
\bibitem [{\citenamefont {Kim}\ and\ \citenamefont
  {Porrati}(2015)}]{kim2015canonical}%
  \BibitemOpen
  \bibfield  {author} {\bibinfo {author} {\bibfnamefont {J.}~\bibnamefont
  {Kim}}\ and\ \bibinfo {author} {\bibfnamefont {M.}~\bibnamefont {Porrati}},\
  }\bibfield  {title} {\bibinfo {title} {{On a canonical quantization of 3D
  Anti de Sitter pure gravity}},\ }\href@noop {} {\bibfield  {journal}
  {\bibinfo  {journal} {Journal of High Energy Physics}\ }\textbf {\bibinfo
  {volume} {2015}},\ \bibinfo {pages} {1} (\bibinfo {year} {2015})}\BibitemShut
  {NoStop}%
\bibitem [{\citenamefont {Cotler}\ and\ \citenamefont
  {Jensen}(2019)}]{cotler2019theory}%
  \BibitemOpen
  \bibfield  {author} {\bibinfo {author} {\bibfnamefont {J.}~\bibnamefont
  {Cotler}}\ and\ \bibinfo {author} {\bibfnamefont {K.}~\bibnamefont
  {Jensen}},\ }\bibfield  {title} {\bibinfo {title} {{A theory of
  reparameterizations for AdS3 gravity}},\ }\href@noop {} {\bibfield  {journal}
  {\bibinfo  {journal} {Journal of High Energy Physics}\ }\textbf {\bibinfo
  {volume} {2019}},\ \bibinfo {pages} {1} (\bibinfo {year} {2019})}\BibitemShut
  {NoStop}%
\bibitem [{\citenamefont {Isham}(1984)}]{isham1984topological}%
  \BibitemOpen
  \bibfield  {author} {\bibinfo {author} {\bibfnamefont {C.~J.}\ \bibnamefont
  {Isham}},\ }\bibfield  {title} {\bibinfo {title} {Topological and global
  aspects of quantum theory},\ }in\ \href@noop {} {\emph {\bibinfo {booktitle}
  {Relativity, groups and topology. 2}}},\ \bibinfo {editor} {edited by\
  \bibinfo {editor} {\bibfnamefont {B.~S.}\ \bibnamefont {DeWitt}}\ and\
  \bibinfo {editor} {\bibfnamefont {R.}~\bibnamefont {Stora}}}\ (\bibinfo
  {publisher} {North-Holland Physics Pub.},\ \bibinfo {year}
  {1984})\BibitemShut {NoStop}%
\bibitem [{\citenamefont {Isham}(1989)}]{isham1989canonical}%
  \BibitemOpen
  \bibfield  {author} {\bibinfo {author} {\bibfnamefont {C.}~\bibnamefont
  {Isham}},\ }\bibfield  {title} {\bibinfo {title} {Canonical groups and the
  quantization of general relativity},\ }\href@noop {} {\bibfield  {journal}
  {\bibinfo  {journal} {Nuclear Physics B-Proceedings Supplements}\ }\textbf
  {\bibinfo {volume} {6}},\ \bibinfo {pages} {349} (\bibinfo {year}
  {1989})}\BibitemShut {NoStop}%
\bibitem [{\citenamefont {Andrade~e Silva}(2022)}]{rodrigo2022}%
  \BibitemOpen
  \bibfield  {author} {\bibinfo {author} {\bibfnamefont {R.}~\bibnamefont
  {Andrade~e Silva}},\ }\bibfield  {title} {\bibinfo {title} {Quantization of
  causal diamonds in 2+1 gravity},\ }\href@noop {} {\bibfield  {journal}
  {\bibinfo  {journal} {{\it In preparation}}\ } (\bibinfo {year}
  {2022})}\BibitemShut {NoStop}%
\bibitem [{\citenamefont {Arnowitt}\ \emph {et~al.}(2008)\citenamefont
  {Arnowitt}, \citenamefont {Deser},\ and\ \citenamefont
  {Misner}}]{arnowitt2008republication}%
  \BibitemOpen
  \bibfield  {author} {\bibinfo {author} {\bibfnamefont {R.}~\bibnamefont
  {Arnowitt}}, \bibinfo {author} {\bibfnamefont {S.}~\bibnamefont {Deser}},\
  and\ \bibinfo {author} {\bibfnamefont {C.~W.}\ \bibnamefont {Misner}},\
  }\bibfield  {title} {\bibinfo {title} {Republication of: The dynamics of
  general relativity},\ }\href@noop {} {\bibfield  {journal} {\bibinfo
  {journal} {General Relativity and Gravitation}\ }\textbf {\bibinfo {volume}
  {40}},\ \bibinfo {pages} {1997} (\bibinfo {year} {2008})}\BibitemShut
  {NoStop}%
\bibitem [{\citenamefont
  {Bartnik}(1988{\natexlab{a}})}]{bartnik1988regularity}%
  \BibitemOpen
  \bibfield  {author} {\bibinfo {author} {\bibfnamefont {R.}~\bibnamefont
  {Bartnik}},\ }\bibfield  {title} {\bibinfo {title} {Regularity of variational
  maximal surfaces},\ }\href@noop {} {\bibfield  {journal} {\bibinfo  {journal}
  {Acta Mathematica}\ }\textbf {\bibinfo {volume} {161}},\ \bibinfo {pages}
  {145} (\bibinfo {year} {1988}{\natexlab{a}})}\BibitemShut {NoStop}%
\bibitem [{\citenamefont {Bartnik}(1988{\natexlab{b}})}]{bartnik1988remarks}%
  \BibitemOpen
  \bibfield  {author} {\bibinfo {author} {\bibfnamefont {R.}~\bibnamefont
  {Bartnik}},\ }\bibfield  {title} {\bibinfo {title} {Remarks on cosmological
  spacetimes and constant mean curvature surfaces},\ }\href@noop {} {\bibfield
  {journal} {\bibinfo  {journal} {Communications in mathematical physics}\
  }\textbf {\bibinfo {volume} {117}},\ \bibinfo {pages} {615} (\bibinfo {year}
  {1988}{\natexlab{b}})}\BibitemShut {NoStop}%
\bibitem [{\citenamefont {Brill}\ and\ \citenamefont
  {Flaherty}(1976)}]{brill1976isolated}%
  \BibitemOpen
  \bibfield  {author} {\bibinfo {author} {\bibfnamefont {D.}~\bibnamefont
  {Brill}}\ and\ \bibinfo {author} {\bibfnamefont {F.}~\bibnamefont
  {Flaherty}},\ }\bibfield  {title} {\bibinfo {title} {Isolated maximal
  surfaces in spacetime},\ }\href@noop {} {\bibfield  {journal} {\bibinfo
  {journal} {Communications in Mathematical Physics}\ }\textbf {\bibinfo
  {volume} {50}},\ \bibinfo {pages} {157} (\bibinfo {year} {1976})}\BibitemShut
  {NoStop}%
\bibitem [{\citenamefont {Gerhardt}(1983)}]{gerhardt1983h}%
  \BibitemOpen
  \bibfield  {author} {\bibinfo {author} {\bibfnamefont {C.}~\bibnamefont
  {Gerhardt}},\ }\bibfield  {title} {\bibinfo {title} {{$H$-surfaces in
  Lorentzian manifolds}},\ }\href@noop {} {\bibfield  {journal} {\bibinfo
  {journal} {Communications in mathematical physics}\ }\textbf {\bibinfo
  {volume} {89}},\ \bibinfo {pages} {523} (\bibinfo {year} {1983})}\BibitemShut
  {NoStop}%
\bibitem [{\citenamefont {Lichnerowicz}(1944)}]{lichnerowicz1944integration}%
  \BibitemOpen
  \bibfield  {author} {\bibinfo {author} {\bibfnamefont {A.}~\bibnamefont
  {Lichnerowicz}},\ }\href@noop {} {\emph {\bibinfo {title} {L'int{\'e}gration
  des {\'e}quations de la gravitation relativiste et le probl{\`e}me des
  n-corps}}}\ (\bibinfo  {publisher} {Gauthier-Villars},\ \bibinfo {year}
  {1944})\BibitemShut {NoStop}%
\bibitem [{\citenamefont {York~Jr}(1972)}]{york1972role}%
  \BibitemOpen
  \bibfield  {author} {\bibinfo {author} {\bibfnamefont {J.~W.}\ \bibnamefont
  {York~Jr}},\ }\bibfield  {title} {\bibinfo {title} {Role of conformal
  three-geometry in the dynamics of gravitation},\ }\href@noop {} {\bibfield
  {journal} {\bibinfo  {journal} {Physical review letters}\ }\textbf {\bibinfo
  {volume} {28}},\ \bibinfo {pages} {1082} (\bibinfo {year}
  {1972})}\BibitemShut {NoStop}%
\bibitem [{\citenamefont {O'Murchadha}\ and\ \citenamefont
  {York~Jr}(1973)}]{o1973existence}%
  \BibitemOpen
  \bibfield  {author} {\bibinfo {author} {\bibfnamefont {N.}~\bibnamefont
  {O'Murchadha}}\ and\ \bibinfo {author} {\bibfnamefont {J.~W.}\ \bibnamefont
  {York~Jr}},\ }\bibfield  {title} {\bibinfo {title} {Existence and uniqueness
  of solutions of the hamiltonian constraint of general relativity on compact
  manifolds},\ }\href@noop {} {\bibfield  {journal} {\bibinfo  {journal}
  {Journal of Mathematical Physics}\ }\textbf {\bibinfo {volume} {14}},\
  \bibinfo {pages} {1551} (\bibinfo {year} {1973})}\BibitemShut {NoStop}%
\bibitem [{\citenamefont {e~Silva}\ and\ \citenamefont
  {Jacobson}(2021)}]{e2021particle}%
  \BibitemOpen
  \bibfield  {author} {\bibinfo {author} {\bibfnamefont {R.~A.}\ \bibnamefont
  {e~Silva}}\ and\ \bibinfo {author} {\bibfnamefont {T.}~\bibnamefont
  {Jacobson}},\ }\bibfield  {title} {\bibinfo {title} {Particle on the sphere:
  group-theoretic quantization in the presence of a magnetic monopole},\
  }\href@noop {} {\bibfield  {journal} {\bibinfo  {journal} {Journal of Physics
  A: Mathematical and Theoretical}\ }\textbf {\bibinfo {volume} {54}},\
  \bibinfo {pages} {235303} (\bibinfo {year} {2021})}\BibitemShut {NoStop}%
\bibitem [{\citenamefont {Witten}(1988{\natexlab{b}})}]{witten1988coadjoint}%
  \BibitemOpen
  \bibfield  {author} {\bibinfo {author} {\bibfnamefont {E.}~\bibnamefont
  {Witten}},\ }\bibfield  {title} {\bibinfo {title} {{Coadjoint orbits of the
  Virasoro group}},\ }\href@noop {} {\bibfield  {journal} {\bibinfo  {journal}
  {Communications in Mathematical Physics}\ }\textbf {\bibinfo {volume}
  {114}},\ \bibinfo {pages} {1} (\bibinfo {year}
  {1988}{\natexlab{b}})}\BibitemShut {NoStop}%
\bibitem [{\citenamefont {Lazutkin}\ and\ \citenamefont
  {Pankratova}(1975)}]{lazutkin1975normal}%
  \BibitemOpen
  \bibfield  {author} {\bibinfo {author} {\bibfnamefont {V.~F.}\ \bibnamefont
  {Lazutkin}}\ and\ \bibinfo {author} {\bibfnamefont {T.~F.}\ \bibnamefont
  {Pankratova}},\ }\bibfield  {title} {\bibinfo {title} {{Normal forms and
  versal deformations for Hill's equation}},\ }\href@noop {} {\bibfield
  {journal} {\bibinfo  {journal} {Functional Analysis and its applications}\
  }\textbf {\bibinfo {volume} {9}},\ \bibinfo {pages} {306} (\bibinfo {year}
  {1975})}\BibitemShut {NoStop}%
\bibitem [{\citenamefont {Segal}(1981)}]{segal1981unitary}%
  \BibitemOpen
  \bibfield  {author} {\bibinfo {author} {\bibfnamefont {G.}~\bibnamefont
  {Segal}},\ }\bibfield  {title} {\bibinfo {title} {Unitary representations of
  some infinite dimensional groups},\ }\href@noop {} {\bibfield  {journal}
  {\bibinfo  {journal} {Communications in Mathematical Physics}\ }\textbf
  {\bibinfo {volume} {80}},\ \bibinfo {pages} {301} (\bibinfo {year}
  {1981})}\BibitemShut {NoStop}%
\bibitem [{\citenamefont {Alekseev}\ and\ \citenamefont
  {Shatashvili}(1989)}]{alekseev1989path}%
  \BibitemOpen
  \bibfield  {author} {\bibinfo {author} {\bibfnamefont {A.}~\bibnamefont
  {Alekseev}}\ and\ \bibinfo {author} {\bibfnamefont {S.}~\bibnamefont
  {Shatashvili}},\ }\bibfield  {title} {\bibinfo {title} {Path integral
  quantization of the coadjoint orbits of the virasoro group and 2-d gravity},\
  }\href@noop {} {\bibfield  {journal} {\bibinfo  {journal} {Nuclear Physics
  B}\ }\textbf {\bibinfo {volume} {323}},\ \bibinfo {pages} {719} (\bibinfo
  {year} {1989})}\BibitemShut {NoStop}%
\bibitem [{\citenamefont {Barnich}\ and\ \citenamefont
  {Compere}(2007)}]{barnich2007classical}%
  \BibitemOpen
  \bibfield  {author} {\bibinfo {author} {\bibfnamefont {G.}~\bibnamefont
  {Barnich}}\ and\ \bibinfo {author} {\bibfnamefont {G.}~\bibnamefont
  {Compere}},\ }\bibfield  {title} {\bibinfo {title} {Classical central
  extension for asymptotic symmetries at null infinity in three spacetime
  dimensions},\ }\href@noop {} {\bibfield  {journal} {\bibinfo  {journal}
  {Classical and Quantum Gravity}\ }\textbf {\bibinfo {volume} {24}},\ \bibinfo
  {pages} {F15} (\bibinfo {year} {2007})}\BibitemShut {NoStop}%
\bibitem [{\citenamefont {Oblak}(2017)}]{oblak2017bms}%
  \BibitemOpen
  \bibfield  {author} {\bibinfo {author} {\bibfnamefont {B.}~\bibnamefont
  {Oblak}},\ }\href@noop {} {\emph {\bibinfo {title} {BMS particles in three
  dimensions}}}\ (\bibinfo  {publisher} {Springer},\ \bibinfo {year}
  {2017})\BibitemShut {NoStop}%
\bibitem [{\citenamefont {Mackey}(1968)}]{mackey1969induced}%
  \BibitemOpen
  \bibfield  {author} {\bibinfo {author} {\bibfnamefont {G.~W.}\ \bibnamefont
  {Mackey}},\ }\href@noop {} {\emph {\bibinfo {title} {{Induced representations
  of groups and quantum mechanics}}}}\ (\bibinfo  {publisher} {Benjamin},\
  \bibinfo {address} {New York, NY},\ \bibinfo {year} {1968})\BibitemShut
  {NoStop}%
\bibitem [{\citenamefont {Mackey}(1963)}]{mackey1963infinite}%
  \BibitemOpen
  \bibfield  {author} {\bibinfo {author} {\bibfnamefont {G.~W.}\ \bibnamefont
  {Mackey}},\ }\bibfield  {title} {\bibinfo {title} {Infinite-dimensional group
  representations},\ }\href@noop {} {\bibfield  {journal} {\bibinfo  {journal}
  {Bulletin of the American Mathematical Society}\ }\textbf {\bibinfo {volume}
  {69}},\ \bibinfo {pages} {628} (\bibinfo {year} {1963})}\BibitemShut
  {NoStop}%
\bibitem [{\citenamefont {Barnich}\ and\ \citenamefont
  {Oblak}(2014)}]{barnich2014notes}%
  \BibitemOpen
  \bibfield  {author} {\bibinfo {author} {\bibfnamefont {G.}~\bibnamefont
  {Barnich}}\ and\ \bibinfo {author} {\bibfnamefont {B.}~\bibnamefont
  {Oblak}},\ }\bibfield  {title} {\bibinfo {title} {{Notes on the BMS group in
  three dimensions: I. Induced representations}},\ }\href@noop {} {\bibfield
  {journal} {\bibinfo  {journal} {Journal of High Energy Physics}\ }\textbf
  {\bibinfo {volume} {2014}},\ \bibinfo {pages} {1} (\bibinfo {year}
  {2014})}\BibitemShut {NoStop}%
\bibitem [{\citenamefont {Barnich}\ and\ \citenamefont
  {Oblak}(2015)}]{barnich2015notes}%
  \BibitemOpen
  \bibfield  {author} {\bibinfo {author} {\bibfnamefont {G.}~\bibnamefont
  {Barnich}}\ and\ \bibinfo {author} {\bibfnamefont {B.}~\bibnamefont
  {Oblak}},\ }\bibfield  {title} {\bibinfo {title} {{Notes on the BMS group in
  three dimensions: II. Coadjoint representation}},\ }\href@noop {} {\bibfield
  {journal} {\bibinfo  {journal} {Journal of High Energy Physics}\ }\textbf
  {\bibinfo {volume} {2015}},\ \bibinfo {pages} {1} (\bibinfo {year}
  {2015})}\BibitemShut {NoStop}%
\bibitem [{\citenamefont {Oblak}(2015)}]{oblak2015characters}%
  \BibitemOpen
  \bibfield  {author} {\bibinfo {author} {\bibfnamefont {B.}~\bibnamefont
  {Oblak}},\ }\bibfield  {title} {\bibinfo {title} {{Characters of the BMS
  group in three dimensions}},\ }\href@noop {} {\bibfield  {journal} {\bibinfo
  {journal} {Communications in Mathematical Physics}\ }\textbf {\bibinfo
  {volume} {340}},\ \bibinfo {pages} {413} (\bibinfo {year}
  {2015})}\BibitemShut {NoStop}%
\bibitem [{\citenamefont {Campoleoni}\ \emph {et~al.}(2016)\citenamefont
  {Campoleoni}, \citenamefont {Gonzalez}, \citenamefont {Oblak},\ and\
  \citenamefont {Riegler}}]{campoleoni2016bms}%
  \BibitemOpen
  \bibfield  {author} {\bibinfo {author} {\bibfnamefont {A.}~\bibnamefont
  {Campoleoni}}, \bibinfo {author} {\bibfnamefont {H.~A.}\ \bibnamefont
  {Gonzalez}}, \bibinfo {author} {\bibfnamefont {B.}~\bibnamefont {Oblak}},\
  and\ \bibinfo {author} {\bibfnamefont {M.}~\bibnamefont {Riegler}},\
  }\bibfield  {title} {\bibinfo {title} {{BMS modules in three dimensions}},\
  }\href@noop {} {\bibfield  {journal} {\bibinfo  {journal} {International
  Journal of Modern Physics A}\ }\textbf {\bibinfo {volume} {31}},\ \bibinfo
  {pages} {1650068} (\bibinfo {year} {2016})}\BibitemShut {NoStop}%
\bibitem [{\citenamefont {Donnelly}\ \emph {et~al.}(2021)\citenamefont
  {Donnelly}, \citenamefont {Freidel}, \citenamefont {Moosavian},\ and\
  \citenamefont {Speranza}}]{donnelly2021gravitational}%
  \BibitemOpen
  \bibfield  {author} {\bibinfo {author} {\bibfnamefont {W.}~\bibnamefont
  {Donnelly}}, \bibinfo {author} {\bibfnamefont {L.}~\bibnamefont {Freidel}},
  \bibinfo {author} {\bibfnamefont {S.~F.}\ \bibnamefont {Moosavian}},\ and\
  \bibinfo {author} {\bibfnamefont {A.~J.}\ \bibnamefont {Speranza}},\
  }\bibfield  {title} {\bibinfo {title} {Gravitational edge modes, coadjoint
  orbits, and hydrodynamics},\ }\href@noop {} {\bibfield  {journal} {\bibinfo
  {journal} {Journal of High Energy Physics}\ }\textbf {\bibinfo {volume}
  {2021}},\ \bibinfo {pages} {1} (\bibinfo {year} {2021})}\BibitemShut
  {NoStop}%
\bibitem [{\citenamefont {Witten}()}]{Edward}%
  \BibitemOpen
  \bibfield  {author} {\bibinfo {author} {\bibfnamefont {E.}~\bibnamefont
  {Witten}},\ }\href@noop {} {}\bibinfo {howpublished} {personal
  communication}\BibitemShut {NoStop}%
\bibitem [{\citenamefont {Achucarro}\ and\ \citenamefont
  {Townsend}(1986)}]{Achucarro:1986uwr}%
  \BibitemOpen
  \bibfield  {author} {\bibinfo {author} {\bibfnamefont {A.}~\bibnamefont
  {Achucarro}}\ and\ \bibinfo {author} {\bibfnamefont {P.~K.}\ \bibnamefont
  {Townsend}},\ }\bibfield  {title} {\bibinfo {title} {{A Chern-Simons Action
  for Three-Dimensional anti-De Sitter Supergravity Theories}},\ }\href@noop {}
  {\bibfield  {journal} {\bibinfo  {journal} {Phys. Lett. B}\ }\textbf
  {\bibinfo {volume} {180}},\ \bibinfo {pages} {89} (\bibinfo {year}
  {1986})}\BibitemShut {NoStop}%
\end{thebibliography}%

\end{document}